\documentclass[manuscript]{acmart}

\usepackage{subcaption}
\usepackage{multirow}

\setcopyright{acmlicensed}
\copyrightyear{2024}
\acmYear{2024}
\acmDOI{XXXXXXX.XXXXXXX}

\acmConference[Conference acronym 'XX]{Make sure to enter the correct
  conference title from your rights confirmation emai}{June 03--05,
  2018}{Woodstock, NY}
\acmISBN{978-1-4503-XXXX-X/18/06}

\author{Emma R. Casolin}
\affiliation{%
  \institution{The University of New South Wales}
  \city{Sydney}
  \state{NSW}
  \country{Australia}}

\author{Flora D. Salim}\authornote{Corresponding author}
\email{flora.salim@unsw.edu.au}
\affiliation{%
  \institution{The University of New South Wales}
  \city{Sydney}
  \state{NSW}
  \country{Australia}}

\author{Ben Newell}
\affiliation{%
  \institution{The University of New South Wales}
  \city{Sydney}
  \state{NSW}
  \country{Australia}}

\begin{document}

%%
%% The "title" command has an optional parameter,
%% allowing the author to define a "short title" to be used in page headers.
\title{Evaluating the Influences of Explanation Style on Human-AI Reliance}

%%
%% The abstract is a short summary of the work to be presented in the
%% article.
\begin{abstract}
  Explainable AI (XAI) aims to support appropriate human-AI reliance by increasing the interpretability of complex model decisions. Despite the proliferation of proposed methods, there is mixed evidence surrounding the effects of different styles of XAI explanations on human-AI reliance. Interpreting these conflicting findings requires an understanding of the individual and combined qualities of different explanation styles that influence appropriate and inappropriate human-AI reliance, and the role of interpretability in this interaction. In this study, we investigate the influences of feature-based, example-based, and combined feature- and example-based XAI methods on human-AI reliance through a two-part experimental study with 274 participants comparing these explanation style conditions. Our findings suggest differences between feature-based and example-based explanation styles beyond interpretability that affect human-AI reliance patterns across differences in individual performance and task complexity. Our work highlights the importance of adapting explanations to their specific users and context over maximising broad interpretability.
\end{abstract}

\begin{CCSXML}
<ccs2012>
   <concept>
       <concept_id>10003120.10003121.10011748</concept_id>
       <concept_desc>Human-centered computing~Empirical studies in HCI</concept_desc>
       <concept_significance>500</concept_significance>
       </concept>
   <concept>
       <concept_id>10010147.10010178</concept_id>
       <concept_desc>Computing methodologies~Artificial intelligence</concept_desc>
       <concept_significance>500</concept_significance>
       </concept>
 </ccs2012>
\end{CCSXML}

\ccsdesc[500]{Human-centered computing~Empirical studies in HCI}
\ccsdesc[500]{Computing methodologies~Artificial intelligence}

\keywords{Explainable AI; Human-AI Collaboration; Human-AI Reliance}

%\received{20 February 2007}
%\received[revised]{12 March 2009}
%\received[accepted]{5 June 2009}

%%
%% This command processes the author and affiliation and title
%% information and builds the first part of the formatted document.
\maketitle

\section{Introduction}\label{ch:intro}

Increasingly, artificial intelligence (AI) is being introduced into human decision-making pipelines in an attempt to support more accurate performance \cite{Overreliance}. However, the growing complexity of AI models prevents sufficient human understanding of the decision processes underlying AI outputs to support their acceptance at face value \cite{XAIReview}. In response to this, explainable AI (XAI) has been proposed as a method of increasing the interpretability of AI models through providing explanations of their behaviour \cite{Transparency}. This increased interpretability seeks to improve human understanding of AI behaviour \cite{SocialSciXAI} and our ability to predict its future behaviour \cite{ProtoCriticism}, with the ultimate goal being that XAI can help humans identify when to appropriately accept and reject AI recommendations \cite{CostBenefit}. In contrast, much of the prior literature has instead found that XAI typically increases people's over-reliance on incorrect AI advice \cite{Reliance1, XAIHelpfulness}, leading to poorer decision outcomes overall.

More recently, evidence has suggested that, in order to support more appropriate human-AI reliance, explanations of AI need to be sufficiently engaged with by users for them to be used to verify AI decisions \cite{CostBenefit}. Thus, a greater understanding of the qualities of explanations that increase engagement with XAI is necessary \cite{Overreliance}. Prior work has found that explanations that are easier to parse are more likely to be engaged with and subsequently used to verify AI decisions \cite{CostBenefit}. Furthermore, the style of explanation, whether it is a feature-based explanation highlighting important features for the model's decision or an example-based explanation displaying examples of similar model inputs and outputs, may also influence how people engage with explanations and the process of verifying the outputs of the AI system generating them \cite{Overreliance}. However, this research has thus far only explored over-reliance, ignoring alternative types of reliance, and has not quantified reliance with respect to uninfluenced human decision-making.

To address these limitations in prior reliance research, models of human-AI reliance that frame the relationship as a casewise difference between human decisions with and without AI support have been proposed \cite{Appropriateness, CausalReliance}. These frameworks allow for the comparison of XAI methods along standardised scales, considering both over- and under-reliance on AI, but have not yet been used to directly compare different styles of explanations. Thus, an understanding of the effects of explanation style on human-AI reliance, including how different styles of XAI may interact in human-AI collaboration scenarios, is missing from the literature. Based on these limitations, we propose the following two research questions to guide our work:

\begin{enumerate}
    \item[\textbf{RQ1:}] Do people exhibit different degrees of appropriate reliance when supported by feature-based compared to example-based explanations, and what are the effects of combining these explanation styles?
    \item[\textbf{RQ2:}] What is the relationship, if any, between interpretability and appropriate human-AI reliance?
\end{enumerate}

Our work addresses these research questions by evaluating the individual and combined influences of feature-based and example-based explanation styles on appropriate human-AI reliance. We measure the impact of these interventions on human-AI decision-making through a two-part, between-subjects randomised experiment exploring reliance and interpretability. Mirroring prior work, \cite{Reliance2, Reliance3, XAIHelpfulness, ConfVsExpln} we do not find evidence towards a positive effect of feature-based, example-based, or combined feature- and example-based explanations on any form of reliance compared to a control condition where AI advice is provided without explanations. Furthermore, we find that explanation styles containing examples are associated with increased reliance on incorrect AI recommendations, despite appearing to improve the interpretability of model predictions. We conclude by contextualising our findings in the literature and exploring opportunities for future work.

\section{Background and Related Work}\label{ch:background}

The complexity of Artificial Intelligence (AI) systems is rapidly increasing to facilitate improved performance and capabilities, often consequently leading to decreased interpretability. This difficulty in understanding the reasoning behind AI model decisions \cite{SocialSciXAI} and predicting its future behaviour \cite{ProtoCriticism} can lead to issues such as the undetected presence of bias \cite{BiasPerception, HumanFramework}, barriers to seeking recourse for people who are negatively impacted by model outputs \cite{Transparency, CFEval}, and the inability to learn new knowledge or techniques from insights made by a model \cite{HumanFramework}.

\subsection{Explanations of AI Systems}

Explainable Artificial Intelligence (XAI) is a collection of methods proposed to improve the transparency and interpretability of AI models. XAI seeks to improve human understanding of AI decision-making by providing explanations of its behaviour and outputs, with the ultimate goal of supporting more appropriate usage of these systems. There are several dimensions across which XAI methods may differ, starting at the level of interpretability at which explanations are generated. For simple, glass-box models, exposed model internals can constitute a complete explanation. This style of direct interpretability includes all possible information regarding the model's decision-making process and is thus always entirely faithful \cite{BlackBoxBad}. However, the large amount of information present in this style of explanation can lead to information overload \cite{InfoOverload} and an overall decrease in understanding \cite{Transparency}, failing to achieve the purpose of XAI.
In contrast to direct interpretability, post-hoc explanations are an alternative approach to XAI that approximates the underlying model. Although this implies a loss of information \cite{Faithfulness}, the resulting simplification can lead to explanations that are easier for humans to understand due to our preference for explanations with fewer causes \cite{CogSciXAI, InterpML}. %Post-hoc XAI methods vary in the amount of information included in their approximations, from model-transparent methods that have full access to the underlying model \cite{Assertiveness} to model-agnostic methods that can only access inputs, outputs, and ground-truth labels \cite{ModelAgnostic}. Post-hoc explanations can also be global or local in scope: global explanations provide an overall understanding of the model that is often a more accurate summary of its behaviour \cite{InterpML}, while local explanations only extend to individual model outputs and are more likely to be inaccurate or inconsistent \cite{XAIReview}. Despite this limitation, local explanations are more effective at increasing understanding of individual model outputs \cite{XAIFairness}, similar to how human explanations are typically produced in response to specific queries \cite{SocialSciXAI}.

XAI methods can additionally be differentiated based on their style of explanation. Two common styles of explanation are feature-based and example-based. Feature-based methods calculate the importance, or contribution, of each feature present in the input towards the final output of the model. Some feature-based methods calculate feature importance scores based on model internals, such as Grad-CAM \cite{GradCAM}, Integrated Gradients \cite{IntegGrad}, and SmoothGrad \cite{SmoothGrad}. Other feature-based methods, such as LIME \cite{LIME} and SHAP \cite{SHAP}, instead generate a surrogate model based on the inputs and outputs of the underlying model from which feature importance scores can be directly observed. In contrast to feature-based methods, example-based XAI methods provide explanations through examples, which can be prototypical, similar, or counterfactual. Prototype explanations generate examples highlighting the prototypical aspects of a class that are important for classification, designed to resemble human reasoning styles \cite{ProtoPNet}. %Conversely, similar and counterfactual example-based explanations produce outputs that are similar to the input being explained, however, counterfactual examples result in a different output from the model \cite{SHAPCounterfact} while similar examples result in the same output from the model as the original input.

\subsection{Human-AI Collaboration}

%Despite the high performance of state-of-the-art AI models, allowing them to act autonomously in high-risk decision-making scenarios can be dangerous, unethical, and illegal \cite{Reliance1}. However, this problem can be mitigated by 
Integrating AI into existing human decision-making pipelines \cite{CausalReliance, Reliance1}, introduces the concept of human-AI collaboration \cite{Overreliance}. Such a process allows an AI agent to make recommendations without removing human accountability over the outcomes of the final decision \cite{Reliance1}. %Ideally, human-AI collaboration should achieve complementary team performance (CTP), defined as when the resulting final decisions are more accurate than either the human or AI alone \cite{Reliance1}. However, achieving a combined accuracy greater than that of an AI system requires that the AI's outputs be verified to decide whether or not they can be relied upon \cite{Reliance4}.
%In response to this need for verification, it has been suggested that increasing human understanding of AI reasoning, such as through the introduction of XAI, may improve human-AI collaboration outcomes \cite{Overreliance}. 
%Contrastingly, 
Prior research \cite{Reliance1, XAIHelpfulness, Reliance2, Reliance3, ConfVsExpln} has found that XAI often leads humans to accept incorrect model predictions without verifying them, typically referred to as over-reliance \cite{CostBenefit, Overreliance}. Prior work \cite{Bias1, Bias2, Bias3, Bias4, Bias5} has suggested that this over-reliance is due to the mere presence of explanations invoking cognitive biases and heuristics that inevitably lead to blind trust in, and deference to, the AI model. However, this theory is flawed in that it contradicts situations in which explanations can be helpful in both human-to-human and human-AI interactions \cite{CostBenefit}.

In contrast to the assumption that over-reliance in human-AI collaboration is inevitable, work by Vasconcelos et al. \cite{CostBenefit} suggests that over-reliance is instead a choice arising out of weighing the costs and benefits of using an explanation to verify the AI. If the cost of engaging with an explanation, in terms of time or cognitive effort, is greater than the cost of blindly following the AI, in terms of the risk of accepting an incorrect prediction, then a rational user would choose not to verify the AI. Similarly, increasing the benefit of a correct final decision, such as through monetary reward, will increase the benefit of using the explanation to verify the AI's prediction, increasing the likelihood of engagement. Over a series of five studies, Vasconcelos et al. found evidence in support of their hypotheses that explanations can reduce over-reliance when they are easier to understand and that people will choose more frequently to engage with explanations when higher performance is incentivised. This finding is further supported by evidence that people are willing to wait longer to receive AI advice when completing more difficult tasks \cite{PsychAI} where the cost of waiting for this support may be perceived as lower than the cost of completing the task without it.
Although this theory can explain why explanations do not improve AI-supported decision-making in all circumstances, some limitations prevent its generalisation to XAI in human-AI collaboration scenarios. %Both the task choice and explanations used in the study were not representative of realistic human-AI collaboration scenarios or XAI methods. The explanations used in their studies contained complete information such that any inconsistencies or errors in the explanation directly implied incorrect reasoning, which was always associated with an incorrect AI prediction. Thus, the explanations could be used as a perfect calibrator for reliance on the AI, which is not true of real XAI methods where the explanations themselves can be a source of unreliability \cite{XAIReview, ImperfectXAI}. %Additionally, due to containing complete information, the explanations could be used by participants to complete the task alone rather than to verify the AI, which may reduce the validity of their findings. Despite these limitations, the presence of any form of cost-benefit evaluation when engaging with AI agents suggests that explanations that are easier for humans to understand, reducing the cost of engaging with them, may reduce over-reliance. Thus, determining which styles of explanations are less costly for humans to engage with may assist with developing XAI that is more likely to be used to verify AI decisions and increase appropriate reliance.

\subsection{Comparing Explanation Styles}\label{sec:chenStudy}

Despite the abundance and variety of existing XAI methods, research comparing the efficacy of different styles of explanations is largely conflicting. Jeyakumar et al. \cite{XAIReview} compared participants' subjective perceptions of the utility of feature-based superimposition-style explanations and explanation-by-example, finding a strong preference for example-based explanations across a variety of input domains provided the examples were visually similar. Wang and Yin \cite{XAIHelpfulness} compared feature-based and example-based explanations on their ability to increase understanding of an AI model, determine the level of uncertainty of the model, and calibrate their reliance on the model, however, they found conflicting results across all three metrics. Du et al. \cite{Reliance11} compared feature-contribution and example-based explanations in a clinical setting, finding no significant differences between these methods on recommendation acceptance rate, while in a similar study using an image classification task, Humer et al. \cite{Reliance12} identified no significant effect of feature-contribution methods and a positive effect of example-based methods for identifying incorrect AI advice only. Without a deeper understanding of how humans engage with and reason about XAI in human-AI collaboration scenarios, these conflicting findings cannot be accounted for.

A holistic understanding of human-AI collaboration with XAI must consider how people interact with both correct and incorrect explanations. Morrison et al. \cite{ImperfectXAI} investigated the influences of feature-based natural language and image example-based XAI methods on human-AI reliance in a bird image classification task, varying the level of participant expertise, assertiveness of AI advice, and correctness of explanations. They found that example-based explanations that disagreed with the AI's suggestions were particularly deceptive, implying that participants may have used the accuracy of the example-based explanations as a simple measure of trust rather than a tool to verify the AI's outputs. However, Morrison et al. refrain from directly comparing the feature-based and example-based explanations in their study due to concerns surrounding the equivalence of information presented between these different explanation styles, particularly due to their different modalities. Additionally, the accuracy rate of the AI in their study was significantly higher than that of many of the participants, making relying on the AI an inherently beneficial strategy overall.

The challenges of measuring and quantifying the effects of AI and XAI support on human-AI collaboration performance, and its relationship to reliance, are further explored in the following section.

\subsection{Measuring and Quantifying Human-AI Reliance}\label{sec:cabitzaMethods}

There is a tendency amongst prior literature to correlate simple measures of accuracy \cite{Overreliance, CostBenefit, Reliance1, Reliance2, Reliance4, Reliance5} or trust \cite{CostBenefit, Reliance1, Reliance2, Reliance3, Reliance4}, based only on a single AI-supported human decision, with reliance. If a human decision-maker accepts an incorrect AI recommendation, this is classified as over-reliance, while rejecting a correct recommendation is classified as under-reliance \cite{CostBenefit}. All other combinations are then classed as ``appropriate reliance'' \cite{CostBenefit}. Under this traditional view of human-AI reliance, summarised in the first column of Figure ~\ref{tab:patterns}, over-reliance, and its less frequently observed counterpart under-reliance \cite{CostBenefit}, can be seen as simply a failure to verify AI decisions appropriately.

\begin{table*}
  \caption{Aligning terminology used in prior research when discussing reliance in human-AI collaboration scenarios. The leftmost three columns state whether the initial human decision, AI decision and final decision are correct (1) or incorrect (0) to produce each reliance pattern, respectively.}
  \label{tab:patterns}
  \begin{tabular}{ccclll}
    \toprule
    Initial & AI & Final & Traditional View \cite{CostBenefit} & Appropriateness of & Technology Dominance \cite{CausalReliance} \\
    Decision & Advice & Decision & & Reliance \cite{Appropriateness} & \\
    \midrule
    0 & 0 & 0 & Over-Reliance & & Detrimental Reliance \\
    0 & 0 & 1 & Appropriate Reliance & & Beneficial Under-Reliance \\
    0 & 1 & 0 & Under-Reliance & Incorrect Self-Reliance & Detrimental Self-Reliance \\
    0 & 1 & 1 & Appropriate Reliance & Correct AI Reliance & Beneficial Over-reliance \\
    1 & 0 & 0 & Over-Reliance & Incorrect AI Reliance & Detrimental Over-Reliance \\
    1 & 0 & 1 & Appropriate Reliance & Correct Self-Reliance & Beneficial Self-Reliance \\
    1 & 1 & 0 & Under-Reliance & & Detrimental Under-Reliance \\
    1 & 1 & 1 & Appropriate Reliance & & Beneficial Reliance \\
    \bottomrule
  \end{tabular}
\end{table*}

\begin{figure}[h]
  \centering
  \includegraphics[width=0.7\linewidth]{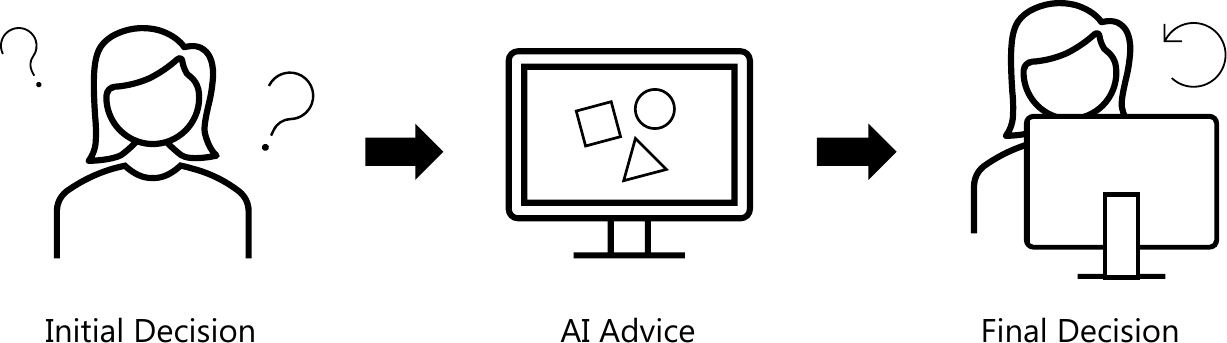}
  \caption{The three stages of the Judge-Advisor System (JAS) Framework.}
  \Description{A diagram with three images connected by arrows pointing from left to right. The first image is of a person surrounded by question marks with the caption ``Initial Decision''. The second image is a computer screen with the caption ``AI Advice''. The third image is of the person from the first image at a computer with the caption ``Final Decision''.}
  \label{fig:jas}
\end{figure}

The main problem with the traditional view of reliance is that, without a measure of human performance on the task both before and after being exposed to an AI intervention, it is impossible to separate true reliance on the intervention from an outcome that may have been observed in its absence \cite{Appropriateness, CausalReliance}. The Appropriateness of Reliance (AoR) \cite{Appropriateness} and Technology Dominance \cite{CausalReliance} views of reliance, summarised in the remaining columns of Figure \ref{tab:patterns}, address the need for a comparison between initial and final human decisions when measuring reliance on AI. This is achieved through a judge-advisor system (JAS) model, shown in Figure \ref{fig:jas}. In this paradigm, the human decision-maker is the judge and the AI is the advisor. By allowing participants to make an initial, unaided decision, it is possible to compare this decision to the final, potentially updated one after AI advice has been received. Both the AoR and Technology Dominance models define quantitative metrics for calculating different types of reliance that can be used to compare AI support interventions.

The AoR framework, proposed by Schemmer et al. \cite{Appropriateness}, was the first to address the need for comparison against baseline human performance in measurements of reliance in human-AI collaboration research. Appropriate reliance under the AoR framework is only defined when CTP is achieved, where the degree of appropriate reliance is defined by the subcomponents Relative AI Reliance (RAIR) and Relative Self-Reliance (RSR). RAIR is the proportion of cases in which humans switch from an initially incorrect decision to agree with a correct AI recommendation on their final decision. The equation for this metric is:
\begin{displaymath}
    \frac{\sum\limits_{i=0}^{N}CAIR_i}{\sum\limits_{i=0}^{N}CA_i}
\end{displaymath}
where for each case $i$, $CAIR_i$ is 1 if the case matches the Correct AI Reliance pathway in Figure \ref{tab:patterns} and 0 otherwise, and $CA_i$ is 1 if the AI recommendation is correct and the human initial decision is incorrect for this case and 0 otherwise. The result is a value between 0 and 1, where higher values indicate more appropriate reliance on AI. Similarly, RSR is the proportion of cases in which humans retain their initial correct decision on their final decision, rejecting an incorrect AI recommendation. The equation for this metric is:
\begin{displaymath}
    \frac{\sum\limits_{i=0}^{N}CSR_i}{\sum\limits_{i=0}^{N}IA_i}
\end{displaymath}
where for each case $i$, $CSR_i$ is 1 if the case matches the Correct Self-Reliance pathway in Figure \ref{tab:patterns} and 0 otherwise, and $IA_i$ is 1 if the AI recommendation is incorrect and the human initial decision is correct for this case and 0 otherwise. The result is once again a value between 0 and 1, where higher values indicate more appropriate reliance on one's initial intuition.

Similar to the AoR model, the Technology Dominance framework, proposed by Cabitza et al. \cite{CausalReliance}, defines metrics based on the same pathways in Figure \ref{tab:patterns}. Instead of proportions, the technology dominance framework uses odds ratios to quantify the degree to which an AI intervention exerts positive or negative dominance over the final decision. These metrics are Detrimental Algorithmic Aversion:
\begin{displaymath}
    \frac{DSR}{N-DSR}\frac{N-BOR}{BOR}
\end{displaymath}
and Automation Bias:
\begin{displaymath}
    \frac{DOR}{N-DOR}\frac{N-BSR}{BSR}
\end{displaymath}
where $N$ is the total number of cases involving the AI intervention and $DSR$, $BOR$, $DOR$, and $BSR$ are the number of cases of Detrimental Self-Reliance, Beneficial Over-Reliance, Detrimental Over-Reliance, and Beneficial Self-Reliance respectively, as defined in Figure \ref{tab:patterns}. Values of the Detrimental Algorithmic Aversion odds ratio less than 1 indicate positive dominance of the AI intervention, where it encourages users to accept correct AI outputs. Higher values of Detrimental Algorithmic Aversion are indicative of lower trust and a reluctance to use AI support to correct a previously incorrect decision. In contrast, values of the Automation Bias odds ratio greater than 1 indicate negative dominance, in which the intervention encourages users to accept incorrect AI outputs. Higher values of Automation Bias are associated with over-trust in an AI intervention, leading people to abandon a previously correct decision in favour of the AI's incorrect prediction.

The AoR and Technology Dominance frameworks, providing standardised, quantitative measures of human-AI reliance, can be used to compare multiple explanation styles across a comparative scale. However, thus far, they have each only been used to compare feature-based XAI to a baseline of AI advice without explanations. Schemmer et al. \cite{Appropriateness} found that the RAIR was higher for participants who were provided with feature-based explanations of AI recommendations compared to those not provided an explanation, however, there was no significant difference in RSR between these conditions, suggesting that the feature-based explanations in this study were only beneficial for helping participants identify and accept correct AI advice. Cabitza et al. \cite{CausalReliance} applied their metrics to four of their prior studies, identifying a higher degree of negative dominance for AI support with explanations compared to without, suggesting a negative effect of feature-based explanations that led participants to over-rely on incorrect AI advice. They found mixed evidence in favour of a difference in positive dominance between the presence and absence of XAI, raising further questions surrounding the relationship between XAI and reliance. It remains unclear whether the patterns of reliance on feature-based explanations observed in these studies extend to other styles of explanations.

\section{Theoretical Development}\label{ch:reliance}

In summary, there has thus far been no research directly comparing different styles of explanations that quantifies human-AI reliance compared to baseline human performance. The result of this is that prior research in this space is difficult to compare, appearing to demonstrate conflicting findings. Chen et al. \cite{Overreliance} observed a disparity between participants' explanation preferences and the explanations that were correlated with higher performance, with participants frequently reporting that feature-based explanations were easier to use and interpret despite displaying higher accuracy with example-based explanations. It remains unclear whether the less intuitive nature of example-based explanations that was reported would prevent sufficient engagement with them for similar performance benefits to be observed in the absence of a think-aloud or other cognitive forcing task. If highlighting features makes explanations more intuitive, and thus more likely to be engaged with, but the nature of example-based explanations and their ability to demonstrate when an AI's prediction is unreliable makes them more useful for calibrating reliance, then combining these aspects of feature-based and example-based explanations may provide additional benefit over either of these explanation styles alone. This theory is supported by evidence that combining explanation modalities can improve human performance in AI-supported decisions \cite{Combinations}. We see the inclusion of a combined explanation style as an important step towards understanding the disparity in performance between feature- and example-based XAI methods under different conditions. 

Another requirement for resolving disparities in the results of prior research is through using a framework with standardised metrics to evaluate reliance. Quantifying reliance through calculated metrics, the Appropriateness of Reliance (AoR) and Technology Dominance models provide frameworks that can be used to compare the results of human-AI collaboration research. Furthermore, their usage of JAS models to separate the effects of an AI intervention from unaided human decision-making isolates true reliance from other relationships in human-AI collaboration. Despite one using proportions and the other using odds ratios, it can be observed that the AoR and Technology Dominance metrics measure the same effects. If these two metrics are calculated for the same dataset, identical plots can be produced by plotting RAIR on the same axis as Detrimental Algorithmic Aversion and RSR on the same axis as Automation Bias, using inverted, logarithmic scales for Technology Dominance. In this unified view, values of the RAIR greater than 0.5 correspond with positive dominance, while values of the RSR less than 0.5 correspond with negative dominance. Due to this equivalence, and the simpler, more easily interpretable calculations used by the AoR metrics, we have chosen to use the AoR framework for the analysis of reliance in our study.

The original RAIR and RSR metrics proposed by Schemmer et al. \cite{Appropriateness} take the sum of all cases together using combined data from all participants. The result is that individual variations in baseline ability on the task between participants can bias the calculations. Specifically, participants with lower baseline accuracy on their initial predictions will skew the RAIR upwards and the RSR downwards, while a participant with higher baseline accuracy will skew the RAIR downwards and the RSR upwards.

\begin{table*}
  \caption{Example responses for two participants completing a task with two cases under the JAS Framework in two different scenarios. Incorrect decisions are marked as 0 while correct decisions are marked as 1.}
  \label{tab:jasexample}
  \begin{tabular}{cccccccc}
    \toprule
    & & \multicolumn{3}{c}{Scenario 1} & \multicolumn{3}{c}{Scenario 2} \\
    & & Initial & AI & Final & Initial & AI & Final \\
    \midrule
    \multirow{2}{6em}{Participant A} & Case 1 & 0 & 1 & 0 & 0 & 1 & 0 \\
     & Case 2 & 1 & 1 & 1 & 1 & 1 & 1 \\
     \multirow{2}{6em}{Participant B} & Case 1 & 0 & 1 & 1 & 0 & 1 & 1 \\
     & Case 2 & 1 & 1 & 1 & 0 & 1 & 1 \\
    \bottomrule
  \end{tabular}
\end{table*}

For example, imagine a task with two cases and two participants, Participant A and Participant B, who perform with the results shown in Scenario 1 of Table \ref{tab:jasexample}. The casewise RAIR for this task would be calculated as $\frac{1}{2}=0.5$, which makes sense as a holistic measurement of Relative AI Reliance for this participant sample on these two cases since Participant A's RAIR is 0 and Participant B's RAIR is 1. Now consider the results from Scenario 2. Again, Participant A's RAIR is 0 and Participant B's RAIR is 1, however now, the casewise RAIR has increased to $\frac{2}{3}=0.67$. Participant B's interactions with the AI have a larger influence on the RAIR calculation in this second example simply due to their lower initial accuracy compared to Participant A, which has the potential to skew the result based on individual differences in technology perception \cite{UserCharacteristics}. For example, people can be hesitant to update their trust in an AI agent due to their prior experiences or presumptions preventing new information from being able to update these prior conceptions \cite{PsychAI}. These individuals may therefore display different reliance patterns due to differences unrelated to experimental manipulations. A similar result can be shown for the RSR if Participant B has a higher initial accuracy compared to Participant A.

To account for this bias, a simple solution is to calculate the RAIR and RSR individually for each participant and take the average of these values. However, a problem arises when using this method in cases where a participant's initial decision is the same as the AI's prediction on every trial, resulting in the RAIR and RSR being evaluated as $\frac{0}{0}$. While it is also possible for this to occur using the casewise versions of these metrics, the probability of this happening for all cases is negligible. In these instances, we evaluate the RAIR and RSR to be 50\%, which is a value representing neither appropriate nor inappropriate reliance. This value reflects the absence of reliance since it implies one is equally likely to rely on themselves and the AI.

\section{Methodology}\label{ch:methods}

\subsection{Chosen Human-AI Collaboration Task}

A widely used task in studies evaluating human-AI collaboration is bird image classification \cite{ProtoPNet, HumanNeeds, DME, SaliencyNLE, BirdsExamples, ImperfectXAI}, as it is a non-trivial task for human participants with a level of difficulty that can be manipulated based on the similarity of selected species. Additionally, bird image classification is a task that AI support is used for in real-world use cases, for example in the Merlin app \cite{HumanNeeds}, and it has been likened to other use cases for AI support, such as medical image diagnosis \cite{ImperfectXAI}. For these reasons, bird image classification is used as the task for participants to complete in our study. Because our participants are not expected to have domain expertise in this area, we provide them with a reference image for each species. This is to prevent a wide discrepancy in human performance, such as that observed by Morrison et al. \cite{ImperfectXAI} in their study using the same dataset.

\subsection{Experimental Design}

Our experimental design consists of two parts, outlined in Figure \ref{fig:protocol}, each utilising different methodologies to measure reliance and interpretability. There are four between-subjects explanation-style conditions:

\begin{figure}[t]
    \caption{Experimental design of our study. Participants are be randomly allocated to one of four explanation-style groups after reading and agreeing to the consent form, remaining assigned to this condition for both Part 1 and Part 2. Before beginning each Part, participants are provided information about task followed by a short quiz to confirm their understanding.}
    \centering
    \includegraphics[width=0.8\linewidth]{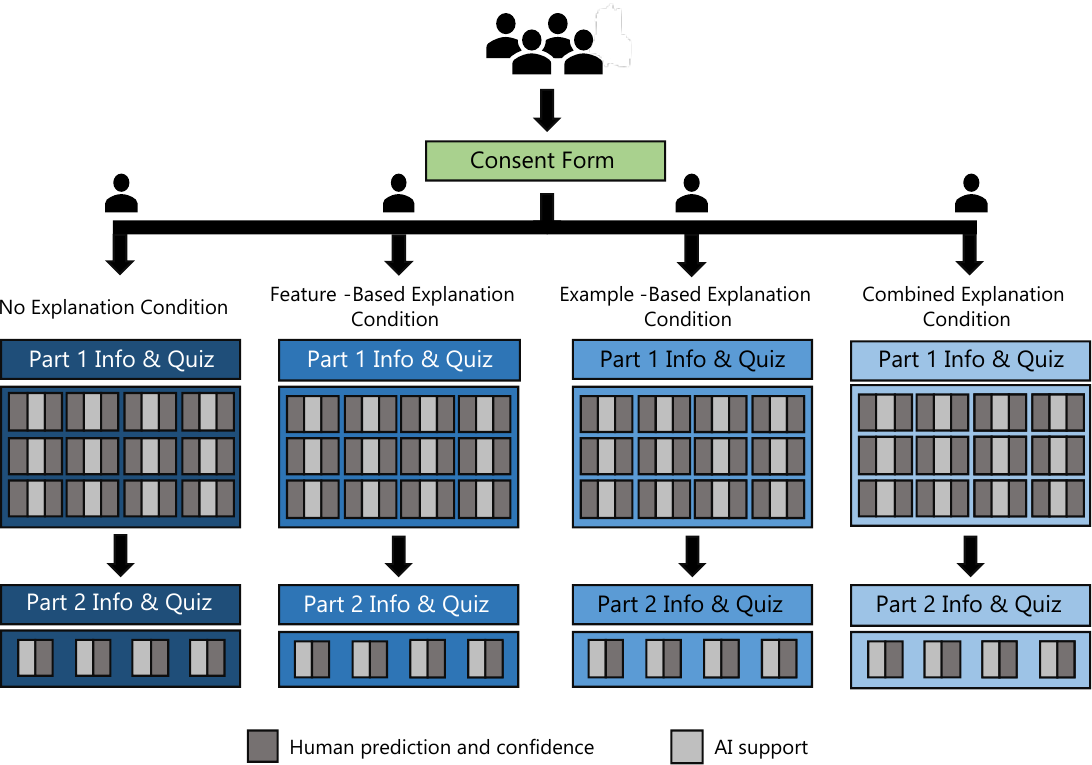}
    \Description{Diagram showing the order of elements in our study. The elements shown in the diagram are described fully in the caption and main text.}
    \label{fig:protocol}
\end{figure}

\begin{enumerate}
    \item The AI prediction without an explanation (control).
    \item The AI prediction supported by a feature-based explanation.
    \item The AI prediction supported by an example-based explanation.
    \item The AI prediction supported by a combined feature- and example-based explanation.
\end{enumerate}

Participants are randomly assigned to one of these conditions at the beginning of the study, remaining in the same condition through Parts 1 and 2. We use a between-subjects design to prevent any potential effects of individual preference between different explanation styles on engagement and subsequent performance. The study is conducted using the online survey platform Qualtrics. Participants provide their informed consent before beginning the study and are debriefed at its conclusion. Before beginning each of the two parts of the study, participants are provided with instructions including a description of how to interpret the explanations if allocated to a condition where the AI provides them. Participants are required to correctly answer questions relating to this information before being allowed to proceed in order to confirm their understanding.

\subsubsection{Part 1: Judge-Advisor System}

In Part 1, we use a judge-advisor system (JAS) model to measure participant reliance on AI advice. There are 12 trials, which we identified as an appropriate number through pilot studies. Each trial comprises three phases: the initial decision, AI support, and final decision. In the first phase, participants see the input case and are asked to name the species of the bird in the image to the best of their ability, with the assistance of the provided reference images. In the second phase, participants are shown the AI's prediction and an explanation, depending on their assigned explanation-style condition. In the final phase, participants can update their final prediction, referring to the original image, their initial decision, and the AI support. In line with the JAS model, we record participants' initial and final decisions on each trial. We additionally ask participants to rate their confidence in the correctness of each answer they give using a validated 5-point Likert scale \cite{Likert}. All participants view the same images across the 12 trials, however, the presentation order is randomised.

Each of the 12 cases presented in Part 1 is assigned a complexity rating from 1 to 4 based on image quality, where low-complexity images (1) show the bird clearly and accurately, low-medium-complexity images (2) are clear but the angle or placement of the bird hides some of its defining features, high-medium-complexity images (3) have the same problems as the previous classification with the addition of some degree of distortion affecting the image, and high-complexity images (4) suffer from issues that severely impact visibility of the bird, such as pixelation or poor lighting that casts the bird in shadow. Participants are not informed of the complexity rating of each bird. Rather, this classification serves to provide additional information for our analysis.

Since individual differences in trust are a confounding factor when measuring reliance \cite{CostBenefit}, we adopt the same controls for mitigating trust as used by Vasconcelos et al. \cite{CostBenefit}. Namely, the AI is not anthropomorphised, its predictions are presented as suggestions that do not need to be accepted, and participants are not informed of their own accuracy until the conclusion of the study to prevent them from using this information to update their perception of the AI's accuracy between trials \cite{CostBenefit}. Additionally, the accuracy rate of the AI is not made known to the participants beyond a statement that the AI's accuracy was similar to that of an ``average person completing this task''. This accuracy rate is set at 58\% (7 out of 12 trials correct) based on the performance of participants in our pilot studies.

% TODO attention checks???

\subsubsection{Part 2: Meta-Predictor Framework}

While Part 1 of our study measures human-AI reliance, it cannot alone provide the necessary insights to explain any differences in reliance between conditions. Part 2 addresses this by including a measure of interpretability based on the meta-predictor framework (MPF) proposed by Colin et al. \cite{HumanFramework}. In Part 2 of our study, participants are required to predict the AI's suggestions on four unseen images using only the provided explanations. Since participants in the no-explanation condition do not have an explanation to assist them, they provide a baseline measure of predictive accuracy that we can compare the explanation conditions against. This part of the study aims to measure how effectively different explanation styles increase interpretability through human understanding of an AI's behaviour \cite{SocialSciXAI} and their ability to predict its future behaviour \cite{ProtoCriticism}. As in Part 1, we ask participants to rate their confidence in their prediction of the AI's suggestion on each trial, and all participants see the same four images in a random order.

In Part 2, there are two within-subjects factors in addition to the between-subjects explanation-style factor. These factors are whether the AI's suggestion for the bird is correct or incorrect and the quality of the provided explanations. We describe how we determine the difference between high- and low-quality explanations in Section \ref{xaimethods}. With four trials in Part 2, each case represents a different combination of levels of the within-subjects factors.

\subsection{Chosen XAI Methods}\label{xaimethods}

In order to isolate the effects of explanation-style from differences in information, an important consideration when selecting the XAI methods for our study is the equivalence of information presented by different generation methods. We note how we address this challenge in this section.

\subsubsection{Example-Based Explanations}

Our example-based explanations are generated using ExMatchina \cite{XAIReview}. ExMatchina constructs explanations by returning the nearest neighbours to the input image from the training data as determined by comparing the cosine similarity of their feature activations at the final convolutional layer. We use the top two returned examples for the explanations in our study, which was the same number used by Chen et al. \cite{Overreliance}. ExMatchina produces higher-quality explanations for larger training datasets, as this increases the probability that the produced nearest neighbours will be more visually similar to the input image \cite{XAIReview}. We considered this factor in our decision to use the CUB dataset for our study, which has approximately 60 training images per class \cite{CUB}. Unlike Chen et al. \cite{Overreliance}, we do not provide ground truth or AI-predicted labels alongside the examples since this could be considered additional information not present in feature-based explanations.

\subsubsection{Feature-Based Explanations}

Our feature-based explanations are generated using SHAP \cite{SHAP}. SHAP is commonly used for generating explanations of AI due to its construction of global explanations that can be applied to local instances, which improves the stability of its explanations across multiple runs for the same input \cite{XAIReview}. SHAP utilises the concept of Shapley values, from game theory, to estimate the contribution of each feature to the final output of the model. For image inputs, the features are taken to be the pixel values. Because ExMatchina is a model-transparent XAI method, we use a model-transparent implementation of SHAP (GradientExplainer) so that our feature-based and example-based explanation styles rely on similar assumptions. This SHAP implementation has also been compared against ExMatchina previously in a study evaluating subjective user preference \cite{XAIReview}.

\subsubsection{Combined Explanations}

\begin{figure}[ht]
    \caption{Equivalence of information as determined through combined explanations between an image (left) and its two nearest-neighbour example-based explanations (middle and right). Ovals of the same colour indicate features shared across the three images.}
    \centering
    \includegraphics[width=\linewidth]{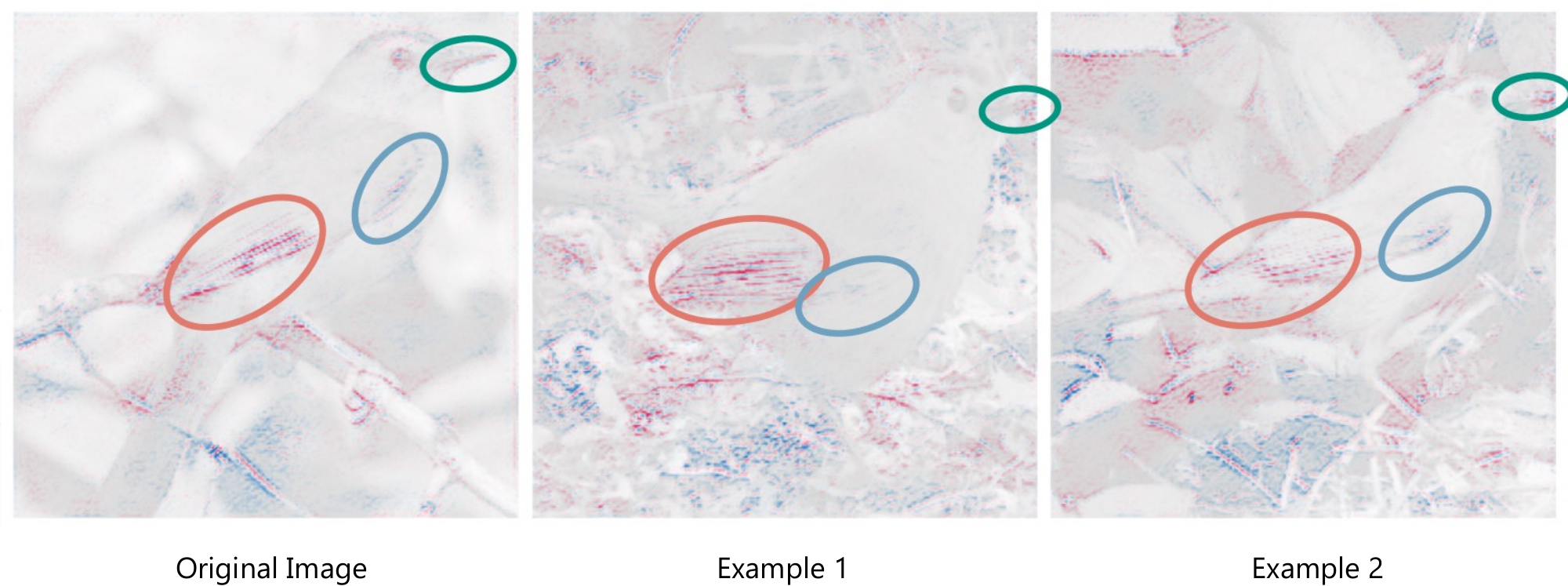}
    \Description{The images are of three similar-looking birds and are in greyscale, with clusters of pixels coloured red or blue by SHAP to indicate features of each image that contributed towards the model's prediciton (red) or against the model's prediction (blue). There are three matching ovals on each image. Oval 1 circles the outlines of each bird's wing feathers in the center of the image, which are highlighted in pink. Oval 2 circles the top of each bird's wing to the right of Oval 1, which is highlighted in blue. Oval 3 circles each bird's beak in the top right corner of the image, which is highlighted in pink.}
    \label{fig:combined}
\end{figure}

Our combined explanations are generated by applying SHAP to the two nearest neighbours determined by ExMatchina. We use the similarity of the SHAP feature highlighting on these examples and the original input image as an approximation for the equivalence of information between the feature-based and example-based explanations. Figure \ref{fig:combined} shows the SHAP feature-based explanations generated for an image and its nearest examples as returned by ExMatchina. From this comparison, we see that the feature-contributions of the three images are highly similar, indicating that these images would be classified based on similar features which are present across all three images. By extension, we argue that the relevant information for determining how the AI classified this image is approximately equivalent between our feature-based, example-based, and combined explanations.

\subsubsection{Explanation Quality}

In Part 2 of our study, one of the variables we explore is how the quality of an explanation affects its utility for interpretability via predictive accuracy. A low-quality explanation that does not completely align with the prediction it is explaining should be less interpretable than a high-quality explanation that perfectly describes the AI's decision. We define a low-quality explanation as one where only one of the examples returned by ExMatchina matches the AI's suggested species, whereas in high-quality explanations both examples match. Because of the assumption we make of the equivalence of information between feature-based and example-based explanations stated previously, we categorise the feature-based and combined explanations associated with the input case for which a low-quality example-based explanation is generated as low-quality as well. While Part 1 of our study does not explicitly consider the effects of explanation quality, we include two images for which low-quality explanations are generated (one each where the AI suggestion is correct and incorrect). This number is too small to form a comparison group, however, it creates a more realistic set of explanations given that XAI is imperfect \cite{ImperfectXAI}.

\subsection{Recruitment}

We use three methods of recruitment to diversify our participant pool. We recruit 95 participants from the crowdsourcing platform Amazon Mechanical Turk (MTurk), 46 participants through convenience sampling in person at the university associated with our research, and 133 participants through the university's mandatory research participation program for undergraduate psychology students. In order to select high-quality participants in our recruitment using MTurk we follow the recruitment strategies used by Colin et al. \cite{HumanFramework}, enforcing that participants of our study have completed at least 10000 tasks and have a task acceptance rate of greater than 98\%. Participants recruited through MTurk receive a base compensation of \$1.20 USD and an additional bonus of \$0.05 USD for each correct final prediction to incentivise engagement with the task and explanations \cite{CostBenefit}. Participants recruited from the university are reimbursed with either a voucher for a free coffee (worth \$5.00) or course credit representative of 15 minutes of research participation.

\subsection{Analysis}\label{sec:analysis}

As discussed in Section \ref{ch:reliance}, our analysis involves using a ``by-participant'' variation of the Appropriateness of Reliance (AoR) framework proposed by Schemmer et al. \cite{Appropriateness}. Our use of the JAS model in Part 1 allows us to calculate the Relative AI Reliance (RAIR) and Relative Self-Reliance (RSR) of participants in each condition. We also use participants' initial and final accuracy to compute complementary team performance (CTP) and the percentage change between initial and final accuracy, which we define as Accuracy Shift. Accuracy Shift factors out any potential improvement in baseline accuracy throughout the task, making it a more direct measure of the ability of participants to utilise AI support when providing their final answers compared to CTP.

Accuracy Shift is calculated for each participant using the equation:
\begin{displaymath}
    \frac{\sum\limits_{n=1}^{12}F_n-I_n}{12}
\end{displaymath}
for each input case $n$, where $F_n$ is 1 if the participant's final decision was correct and 0 otherwise and $I_n$ is 1 if the participant's initial decision was correct and 0 otherwise. We use a similar formula to calculate the change in participant confidence between with initial and final decisions. In Part 2 we only analyse participants' predictive accuracy as a binary value for each case dependent upon whether or not they correctly predict the AI's suggestion.

Due to the conflicting findings of prior work discussed in Chapter \ref{ch:background}, our research is exploratory in nature, thus utilising a post-hoc analysis strategy. Calculating the AoR metrics per each participant allows us to conduct an analysis based on the spread of the distribution of data across all participants, or analyses of variance. Although surveys have been used extensively in prior human-centered XAI research to analyse the subjective feelings of participants towards the explanations and their own self-perceived performance or confidence, we limit our analysis of subjective data due to its poor reliability and high susceptibility to demand characteristics. Such subjective measures additionally do not tend to align with objective measures in XAI research that utilise them \cite{Overreliance, HumanExpln, XAIHelpfulness, Reliance9}.
\section{Results}

We begin our analysis with the results from the JAS trials in Part 1, measuring human-AI reliance. The descriptive results from this part of the study are presented in Table \ref{tab:part1}. Results are reported as confidence intervals (CIs) for comparison differences in dependent variable units.

\begin{table*}
  \caption{Descriptive results from Part 1 of the study}
  \label{tab:part1}
  \begin{tabular}{lcccccc}
    \toprule
    Condition & Initial & Final & Accuracy & Confidence & RAIR (SD) & RSR (SD) \\
    & Accuracy (SD) & Accuracy (SD) & Shift (SD) & Shift (SD) & & \\
    \midrule
    No Explanation & 59.7\% (19.0\%) & 66.9\% (19.3\%) & 7.2\% (10.7\%) & 0.21 (0.75) & 41.1\% (37.3\%) & 88.8\% (24.1\%) \\
    Feature-Based & 61.4\% (20.9\%) & 65.5\% (17.7\%) & 4.1\% (10.1\%) & 0.18 (0.75) & 33.1\% (30.7\%) & 83.7\% (23.6\%) \\
    Example-Based & 58.1\% (18.5\%) & 65.4\% (16.1\%) & 7.3\% (13.3\%) & 0.19 (0.78) & 49.0\% (29.7\%) & 79.3\% (29.9\%) \\
    Combined & 57.8\% (13.5\%) & 63.4\% (12.8\%) & 5.5\% (11.3\%) & 0.18 (0.83) & 50.4\% (33.2\%) & 72.7\% (33.0\%) \\
    \bottomrule
  \end{tabular}
\end{table*}

\subsection{Complementary Team Performance}

\begin{figure}[ht]
    \centering
    \includegraphics[width=0.7\linewidth]{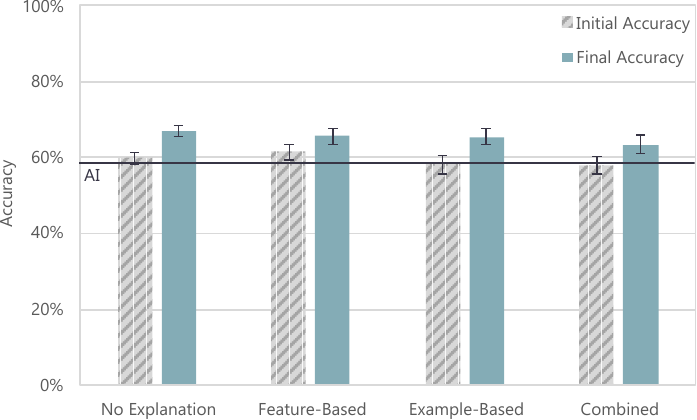}
    \caption{Participants' average initial and final accuracy per experimental condition, relative to the accuracy of the AI.}
    \label{fig:ctp}
\end{figure}

Complementary Team Performance (CTP) measures the influence of an AI intervention on human-AI team decision-making compared to baseline human and AI performance individually. Using a Bonferroni-t analysis, we find that participants are more accurate in their final decisions than their initial ones across all conditions (No Explanation 99\% CI: 2.6-11.8\%; Feature-Based 95\% CI: 0.4-7.8\%; Example-Based 99\% CI: 2.8-11.7\%; Combined 99\% CI: 1.0-10.0\%). Similarly, we use Bonferroni-adjusted one-sample t-tests to compare final accuracy against the AI's accuracy of 58\%, finding a positive, significant difference for all conditions (No Explanation 99\% CI: 4.5-12.7\%; Feature-Based 99\% CI: 2.3-12.1\%; Example-Based 99\% CI: 1.6-12.4\%; Combined 95\% CI: 0.6-9.5\%). Thus, we can conclude that CTP is achieved by participants provided with any of the three forms of XAI included in this study and those without XAI support (Figure \ref{fig:ctp}).

\begin{figure}[ht]
    \centering
    \includegraphics[width=0.7\linewidth]{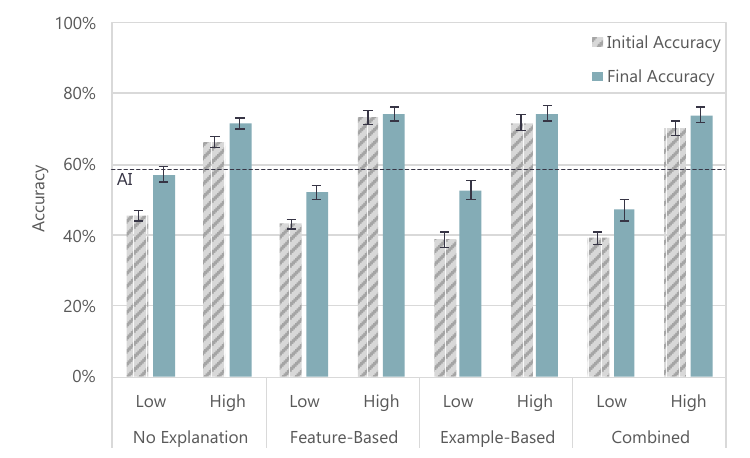}
    \caption{Participants' average initial and final accuracy per experimental condition split by task ability, relative to the accuracy of the AI. Error bars show standard error of the mean.}
    \label{fig:ctpsplit}
\end{figure}

In the calculation of CTP in Figure \ref{fig:ctp}, responses from all participants in each condition are averaged, masking individual performance relative to the AI. CTP proposes to measure the effect of AI support irrespective of differences in performance between humans and AI \cite{Reliance1}, however, by averaging across all participants, the ability to compare the effect of AI support on final accuracy between individuals who initially perform above or below the AI's level of accuracy is lost. In Figure \ref{fig:ctpsplit} we separate participants in each condition by their baseline ability on the task, where those classified as having ``low ability'' have an average score across all trials that is less than the AI's accuracy of 58.3\% and those classified as having ``high ability'' perform with an average accuracy greater than or equal to this value. Once participant ability on the task is accounted for, we see that an improvement in final accuracy over initial accuracy is seen in all low-ability participants (No-Explanation 99\% CI: 1.9-21.1\%; Feature-Based 99\% CI: 0.6-17.3\%; Example-Based 99\% CI: 5.6-22.0\%; Combined 95\% CI: 1.4-14.7\%) and for high-ability participants in the No-Explanation condition (90\% CI: 0.2-10.0\%). However, because a positive significant difference between AI and final accuracy is only observed in high-ability participants (No Explanation 99\% CI: 7.4-19.1\%; Feature-Based 99\% CI: 9.3-22.5\%; Example-Based 99\% CI: 8.5-23.6\%; Combined 99\% CI: 8.6-22.7\%), we see that the requirements for CTP are only achieved by high-ability participants not provided with an explanation, and the significance of this finding is weak.

\subsection{Accuracy Shift}

\begin{figure}[ht]
     \centering
     \begin{subfigure}{0.45\textwidth}
     \centering
         \includegraphics[width=\textwidth]{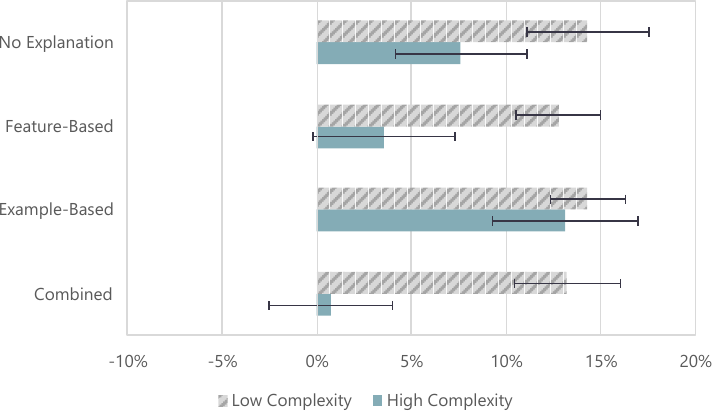}
         \caption{Low-Ability}
         \label{fig:accshift_low}
     \end{subfigure}%
     \quad
     \begin{subfigure}{0.45\textwidth}
     \centering
         \includegraphics[width=\textwidth]{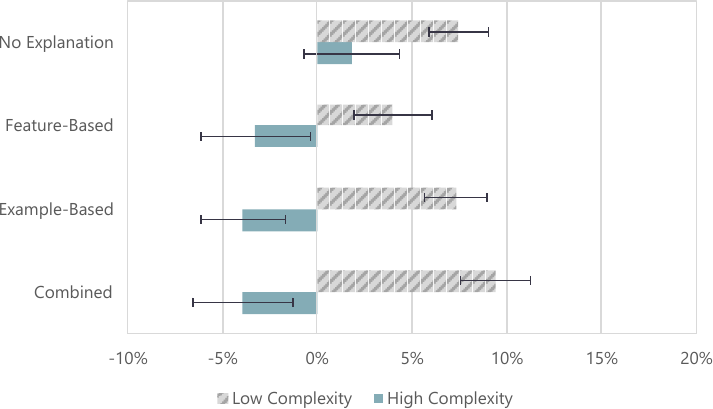}
         \caption{High-Ability}
         \label{fig:accshift_high}
     \end{subfigure}
     \caption{Accuracy Shift on low- and high-complexity cases for participants with low (a) and high (b) ability on the task. Error bars show standard error of the mean.}
     \label{fig:accshift}
\end{figure}

Following the observed differences in CTP between participants of differing ability on the task, we compare the Accuracy Shift of high- and low-ability participants across explanation-style conditions and task complexities in Figure \ref{fig:accshift}. We group low- and low-medium-complexity images together into a single low-complexity group, and medium-high- and high-complexity images together as high-complexity based on our preliminary observation that there was little difference in participant performance within these pairs.

We analyse this data through a post-hoc Scheff\'e contrasts analysis. Accuracy Shift is higher for low- compared to high-complexity images across all conditions for both low-ability (99\% CI: 13.1-78.0\%) and high-ability (99\% CI: 5.3-13.4\%) participants. We see little difference in Accuracy Shift between conditions on low-complexity cases, while greater variability in performance can be observed for high-complexity cases. In the case of low-performing participants (Figure \ref{fig:accshift_low}), the greatest improvement in accuracy is seen for those provided with example-based explanations, while little improvement is seen for feature-based or combined explanations. For higher-performing participants (Figure \ref{fig:accshift_high}), all forms of XAI are associated with a worsening of accuracy, whereas having no explanation appears to lead to improvement. Although these comparisons are interesting to consider, they are not statistically significant.

\subsection{Appropriateness of Reliance}

\begin{figure}[ht]
    \centering
    \includegraphics[width=0.5\textwidth]{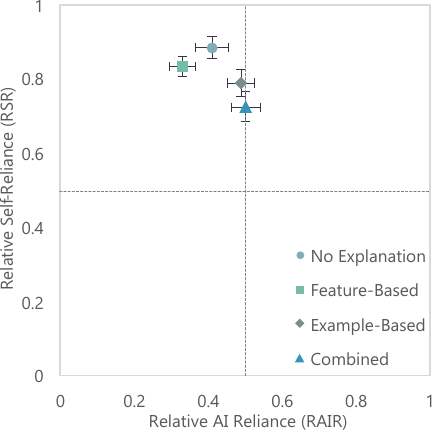}
    \caption{Appropriateness of reliance across explanation styles. Error bars show standard error of the mean.}
    \label{fig:aor}
\end{figure}

We now explore differences in human-AI reliance across explanation styles through Relative AI Reliance (RAIR) and Relative Self-Reliance (RSR). We begin by conducting separate ANOVAs for each of these two metrics to test whether there are differences between conditions, finding both to be significant (RAIR: $F=4.152, p=0.007$; RSR: $F=4.016, p=0.008$). Thus, we can continue with a post-hoc Scheff\'e contrasts analysis to further explore comparisons between explanation styles. The findings from this analysis are shown in Figure \ref{fig:aor}.

For the RAIR, we find a strong significant difference between feature-based explanations and any explanation style with examples (example-based and combined explanations), where the RAIR is 0.3-32.9\% higher for explanations with examples (99\% CI). This comparison is also significant when combining the no-explanation group with the feature-based explanation group, for which the 95\% CI for the difference is 1.5-23.7\%. We conduct Bonferroni-adjusted one-sample t-tests to test for the presence or absence of positive dominance \cite{CausalReliance}, which we take to be values of the RAIR greater or less than 50\%, respectively. From this analysis, we find that feature-based explanations ($t=-4.639$, $p=0.000$) and having no explanation ($t=-1.934$, $p=0.057$) each do not exert positive dominance. There is insufficient evidence to determine whether example-based explanations ($t=-0.295$, $p=0.769$) or combined explanations ($t=0.097$, $p=0.923$) exert positive dominance or not.

In contrast to the RAIR, for the RSR we see poorer reliance when participants are provided explanations with examples compared to without. For participants who receive explanations with examples, their RSR is 1.0-24.6\% lower compared to those who receive no explanations (95\% CI) and 0.8-19.8\% lower when including participants provided with feature-based explanations alongside those who do not receive an explanation (95\% CI). Again, we conduct Bonferroni-adjusted one-sample t-tests, this time to test for the presence or absence of negative dominance \cite{CausalReliance}, which we define as the RSR being less than or greater than 50\%, respectively. Across all four conditions, we find that negative dominance is not exerted (No Explanation: $t=12.979$, $p=0.000$; Feature-Based: $t=-12.024$, $p=0.000$; Example-Based: $t=8.208$, $p=0.000$; Combined: $t=5.674$, $p=0.000$).

Because we saw an effect of participant task ability and case complexity on Accuracy Shift, we similarly analyse AoR across these factors, performing separate post-hoc Scheff\'e analyses for high- and low-ability participants. We find a statistically significant, but weak, improvement in RAIR when high-ability participants receive explanations with examples compared to those who receive feature-based explanations (90\% CI: 1.6-28.6\%), which is not present for low-ability participants. The opposite is seen for the RSR of high-ability participants, where there is a weak decrease between participants who receive feature-based explanations compared to those who receive an explanation containing examples (90\% CI: 0.9-21.8\%). We also observe that, averaged across conditions, high-ability participants display higher RAIR (95\% CI: 1.2-15.8\%) and RSR (95\% CI: 0.3-12.1\%) on high- compared to low-complexity cases.

\subsection{Interpretability}

Interpretability is explored in Part 2 of the study. We conduct a post-hoc Scheff\'e contrasts analysis to test the significance of simple effects and interactions between explanation quality (Quality), AI correctness (Correctness), and explanation style (Condition). All of the following results are provided with 95\% simultaneous confidence interval estimates. Beginning with the interaction between Quality and Condition, high-quality explanations result in 25.5\% ($\pm 23.0\%$) better predictive accuracy than low-quality explanations for explanations with examples (i.e. example-based and combined explanations) compared to feature-based explanations. Furthermore, when the AI's prediction is incorrect this difference increases to 33.1\% ($\pm 31.6\%$). Overall, these findings suggest that the quality of XAI explanations is important when considering interpretability, particularly for explanations provided through examples.

% \begin{figure}[ht]
%      \centering
%      \begin{subfigure}[b]{0.49\textwidth}
%          \centering
%          \includegraphics[width=\textwidth]{charts/mpf/manova/qual.pdf}
%          \caption{All AI}
%          \label{fig:qual}
%      \end{subfigure}
%      \hfill
%      \begin{subfigure}[b]{0.49\textwidth}
%          \centering
%          \includegraphics[width=\textwidth]{charts/mpf/manova/qualwrong.pdf}
%          \caption{AI Incorrect}
%          \label{fig:qualwrong}
%      \end{subfigure}
%         \caption{Interactions Between Explanation Quality and Explanation Style}
%         \label{fig:qualeffects}
% \end{figure}

For the interaction between Correctness and Condition, we find that predictive accuracy is lower on cases where the AI's prediction is incorrect compared to when it is correct by 10.5\% ($\pm 7.0\%$). This effect is stronger when a feature-based explanation is provided compared to when an explanation with examples is provided by 37.2\% ($\pm 24.8\%$). Additionally, we find that this effect of AI correctness is enhanced for high-quality explanations, for which the difference in predictive accuracy between correct and incorrect AI suggestions is 44.8\% ($\pm 30.4\%$) wider for participants given feature-based explanations compared to those given explanations with examples. Taken together, this suggests that correct AI suggestions are more predictable than incorrect ones, but that explanations through examples can help humans more accurately identify the predictions made by an unreliable AI agent.

% \begin{figure}[ht]
%      \centering
%      \begin{subfigure}[b]{0.49\textwidth}
%          \centering
%          \includegraphics[width=\textwidth]{charts/mpf/manova/correctness.pdf}
%          \caption{Any Quality}
%          \label{fig:corr}
%      \end{subfigure}
%      \hfill
%      \begin{subfigure}[b]{0.49\textwidth}
%          \centering
%          \includegraphics[width=\textwidth]{charts/mpf/manova/corrhighqual.pdf}
%          \caption{High Quality Explanations}
%          \label{fig:corrhighqual}
%      \end{subfigure}
%         \caption{Interactions Between AI Correctness and Explanation Style}
%         \label{fig:correffects}
% \end{figure}

We finish by highlighting significant simple effects identified in our analysis. First, when the AI's prediction is incorrect, humans are 33.6\% ($\pm 21.0\%$) better at guessing its prediction when provided with high-quality explanations with examples compared to high-quality feature-based explanations. Second, when the AI's prediction is correct, humans are 29.1\% ($\pm 23.9\%$) worse at guessing its prediction when provided with low-quality explanations with examples compared to low-quality feature-based explanations. Overall, we find that the effect of explanation style on AI interpretability is mediated by both the accuracy of AI and the quality of the provided explanations.

% \begin{figure}[ht]
%      \centering
%      \begin{subfigure}[b]{0.49\textwidth}
%          \centering
%          \includegraphics[width=\textwidth]{charts/mpf/manova/lowcorrect.pdf}
%          \caption{Low Quality Explanations of Correct AI Predictions}
%          \label{fig:lowcorrect}
%      \end{subfigure}
%      \hfill
%      \begin{subfigure}[b]{0.49\textwidth}
%          \centering
%          \includegraphics[width=\textwidth]{charts/mpf/manova/highincorrect.pdf}
%          \caption{High Quality Explanations of Incorrect AI Predictions}
%          \label{fig:highincorrect}
%      \end{subfigure}
%         \caption{Simple Effects of Explanation Style on Predictive Accuracy}
%         \label{fig:simpleeffects}
% \end{figure}

% \subsection{Summary}

% The results of our meta-prediction task find that predictive accuracy is only improved above baseline by explanation styles containing examples. This result is surprising considering that example-based and combined explanations were the least supportive of appropriate reliance in Part 1 of our study, suggesting that greater interpretability of AI decisions does not necessarily promote better utility. Notably, the increased ability of explanations with examples to help participants correctly determine the AI's prediction when it was correct was primarily isolated to cases where explanations were of a low quality.
\section{Discussion}

Our results explore the influence of explanation style on performance in human-AI collaboration tasks. We compare feature-based, example-based, and combined feature- and example-based explanations against a baseline control condition of AI suggestions without explanations to answer our research questions. We measure the effects of explanation style on reliance in the first part of our study, addressing RQ1, and on interpretability in the second part, exploring RQ2. In this section, we summarise and contextualise the main findings of this study.

\textbf{Our findings indicate that humans display more appropriate AI reliance when provided XAI containing examples compared to feature-based XAI.} In our analysis, we observe lower values of RAIR for feature-based compared to example-based and combined explanations, which is significant at the 0.01 level. This suggests that humans are better at using information about correct AI reasoning to calibrate their reliance on an AI agent when this information is presented through examples, as opposed to feature highlighting. This is further supported by our finding that explanation quality has a greater effect on AI interpretability for explanations provided through examples compared to feature-based ones, implying that incorporating examples into explanations of AI is particularly beneficial for appropriate trust calibration on reliable AI systems, provided the explanations are of good quality.

Although we observe lower values of RAIR in participants who receive feature-based compared to no explanations, this difference is not statistically significant. This aligns with the results of Cabitza et al. \cite{CausalReliance}, who reported negligibly poorer reliance on correct AI advice for participants shown feature-based explanations compared to those who were not. Contrastingly, Schemmer et al. \cite{Appropriateness} observed the opposite result in their study, where higher values of RAIR were displayed by participants receiving feature-based explanations compared to no XAI. A difference between the studies in these papers is that, while those analysed by Cabitza et al. featured multiclass image classification tasks similar to ours, the study conducted by Schemmer et al. involved binary text classification instead. There may be different effects of explanation style across task modalities and domain magnitude, which requires further exploration.

% These findings are consistent with those of Chen et al. \cite{Overreliance}, who found no significant difference in accuracy between feature-based and example-based explanations when the AI was correct. Similarly, Morrison et al. \cite{ImperfectXAI} also observed similar RAIR values for feature-based and example-based explanations. Furthermore, we were unable to find evidence of CTP across any of the three explanation conditions, which was mirrored in our analysis of reliance where XAI did not improve RAIR or RSR relative to AI suggestions without explanations.  An interesting finding from our work was that, despite being the only condition to exhibit CTP, the Accuracy Shift, RAIR, and RSR were all found to display a decreasing trend over time. 

\textbf{We find that humans exhibit less appropriate self-reliance when provided explanations containing examples compared to the absence of examples.} The natural consequence of the increased effect of explanation quality for example-based explanations on interpretability and subsequent reliance is that high-quality example-based explanations can be deceiving. In our study, we observe a significantly lower RSR of participants who are provided with any form of XAI containing examples compared to those who are not. This result remains significant when comparing XAI containing examples with AI suggestions without explanations, but not feature-based XAI alone. Our findings align with several studies in the literature. For example, despite seeing minimal difference between the RSR of participants provided with feature-based compared to example-based explanations, Morrison et al. \cite{ImperfectXAI} noted that domain experts were more susceptible to being misled by correct example-based explanations of incorrect AI recommendations. Because the majority of the explanations used in our study aligned with their definition of correct explanations (which is similar to our definition of high-quality explanations), this could explain why our participants, who had reference images to assist them, were similarly misled. Similar to our findings, Wang and Yin \cite{XAIHelpfulness} found that feature-based explanations helped individuals accept correct AI predictions and reject incorrect ones, but this was not true for example-based explanations. In the context of these findings, our results suggest that example-based explanations do not support humans' ability to correctly verify AI outputs, but rather, they can lead to overconfidence in incorrect AI suggestions. This is particularly concerning when considered alongside our finding that people are better at identifying incorrect AI predictions through the information provided in examples compared to feature highlighting. This creates the potential for high-quality example-based explanations to mislead users into accepting unreliable AI recommendations, which was the outcome observed by Morrison et al. in their study \cite{ImperfectXAI}.

% Similar to the results of prior studies \cite{Appropriateness, CausalReliance, Reliance1, Reliance2}, we find no significant differences between the detrimental reliance of participants provided with feature-based explanations compared to those who received AI advice without explanations. In line with our own results, reliance was on the beneficial side of the scale for both measures. However, in their study, Chen et al. \cite{Overreliance} observed a decrease in accuracy when the AI gave an incorrect recommendation with feature-based explanations, but not with example-based explanations. Additionally, several studies in the past have concluded that feature-based XAI similarly results in over-reliance on AI \cite{Reliance1, XAIHelpfulness, Overreliance}. However, we note that these studies either did not compare their results to a control condition where AI advice was provided without an explanation \cite{Overreliance}, or found no difference between this control condition and feature-based explanations \cite{Reliance1, Reliance2}. Combined, this suggests the presence of additional factors that influence human engagement with AI, regardless of XAI.

% However, our findings diverge from those of Humer et al. \cite{Reliance12} and Chen et al. \cite{Overreliance}, who reported no significant effects for feature-based explanations but found a positive effect for example-based explanations when the AI was incorrect. 

\textbf{Combined feature- and example-based explanations do not support better human-AI reliance compared to feature-based and example-based explanation styles alone.} Overall, we find insufficient evidence to suggest that the RAIR or RSR of participants supported through combined explanations differ from those who receive example-based explanations alone. Similarly, differences between the reliance of participants provided with combined and feature-based explanations are only of statistical significance when combined explanations are grouped with example-based explanations in analyses. Taken together, we conclude that combined feature- and example-based explanations are no more beneficial or detrimental for human-AI reliance than the component explanations they subsume. This finding somewhat contradicts those of Robbemond et al. \cite{Combinations}, who found that explanations that combined different presentation modalities were more beneficial for decision accuracy. However, since the combined explanations in our study involve two visual forms of explanation, it may be the case that combining explanations of the same modality simply increases the amount of information available, leading to information overload \cite{InfoOverload} rather than improved understanding. Because the combination of explanation styles in our study is designed to present repeated information in different ways rather than additional, new information, this heightens the risk of providing unnecessary information that may become overwhelming to users.

\textbf{Baseline human performance and case complexity can influence human-AI reliance.} We observe differences between both high- and low-ability participants and high- and low-complexity cases, aligning with research suggesting that XAI is more effective when it caters to its users and context \cite{AdaptingXAI}. In our analysis of RAIR and RSR across participant abilities, we find that the differences in reliance across explanation styles seen when averaging across all participants are only present in the high-performing group. This group also displays more appropriate reliance across all conditions on high- compared to low-complexity cases. This is in contrast to our observation that people with low ability on the task show a greater improvement in accuracy between their initial and final answers, which is particularly noticeable on high-complexity cases, as seen in our analysis of Accuracy Shift. This discrepancy highlights the difference between reliance and changes in decision accuracy, supporting the notion that these are two separate concepts that cannot be used interchangeably \cite{Appropriateness}. Improvements in the performance of low-ability participants in our study are solely attributable to the support of AI recommendations more accurate than their own predictions \cite{Reliance1}. In contrast, despite showing minimal improvement in accuracy, high-ability participants demonstrate distinct patterns of reliance that respond to task complexity and explanation style. If the higher baseline performance of these participants was due to a higher degree of engagement with the task, and by extension the explanations, our findings also provide support for the role of engagement in the cost-benefit theory of human-AI reliance \cite{CostBenefit}.

\textbf{Increased interpretability does not necessarily promote better human-AI collaboration.} Despite example-based and combined explanations being more helpful for determining the recommendation made by an AI agent when it is incorrect or when the explanations are of high quality, this does not correspond with more appropriate reliance. Because almost all of the explanations in Part 1 meet our definition of high-quality explanations, we would expect both the RAIR and RSR of participants who receive explanations containing examples to be higher than those who do not if interpretability improves reliance, however, this is seen only for the RAIR. Additionally, we observe that explanations containing examples only help participants predict the AI's suggestion when the examples appear significantly different from the original image, as is the case for the high-quality explanation of an incorrect AI suggestion. This raises the question of whether our experimental design in Part 2 truly measures interpretability; although participants could determine \textit{what} the AI was predicting, this does not mean they were able to determine \textit{why} it made that prediction.
\section{Limitations and Future Work}

A methodological limitation of our study is the inclusion of only one method of feature-based XAI and one method of example-based XAI. We make the simplifying assumption that SHAP represents all types of feature-based and ExMatchina represents all types of example-based explanations, however, this assumption is unlikely to be true in practice and limits the generalisability of our findings. Further research should examine and compare alternative XAI methods within and beyond feature-based, example-based, and combined explanation styles. The consideration of additional input and explanation modalities, such as textual inputs and natural language explanations, is also an avenue for future work, particularly in the context of different human-AI collaboration tasks.

The explanations used in our study are additionally constrained by a lack of control over the amount of information presented. The combined explanations contain a large amount of information, which likely resulted in some amount of information overload \cite{InfoOverload}. Simplified feature highlighting or the application of image augmentation techniques to the examples to more clearly emphasise feature similarities in combined explanations may help to reduce this effect. Furthermore, despite controls for the equivalence of useful information between feature-based and example-based explanations, there is a possibility that the additional uninformative information contained in image examples may influence participant engagement with example-based and combined explanations \cite{HumanFramework}. The inclusion of additional control conditions for uninformative or incorrect explanations may help to further clarify this relationship. 

Finally, as has been mentioned in prior work, there is still a need to explore human-AI reliance beyond the cases where an AI intervention makes a suggestion that differs from the initial human decision \cite{Appropriateness}. Adding to this, we note that the relationship between human decision-making and AI support becomes more complex when extended beyond binary decisions, such as with the multiclass classification task used in this study. The reliance frameworks explored throughout this paper consider the correctness of a human or AI prediction through a binary lens, however, if there are multiple correct or incorrect answers, or if there is a hierarchy to the correctness of answer choices, then switching to or from an AI's suggestion has a wider range of potential consequences. Additionally, in real-life decisions, there is often more to consider than the correctness of an outcome, such as the risks and implications of accepting it. Future XAI reliance research should consider scenarios that require maximising several goals to more closely replicate the stakes of real-world decisions.
\section{Conclusion}\label{ch:conclusion}

Adding to the growing body of research investigating the effectiveness of XAI for improving human-AI collaboration, our work evaluated the influence of two different explanation styles, feature-based and example-based, on human-AI reliance patterns. Additionally, we investigated the effects of combining these explanation styles to provide a better understanding of how different forms of XAI may complement each other to increase user understanding. We conducted a two-part study exploring reliance and interpretability across these three explanation styles, comparing them to a control condition where AI advice was provided without explanations. The findings of our research contribute towards recognising and improving the qualities of XAI that promote more positive human-AI collaboration outcomes.

\bibliographystyle{ACM-Reference-Format}
\bibliography{pubs}

\end{document}